# Single-channel EEG features during n-back task correlate with working memory load

Neta B. Maimon[1,2], Lior Molcho[2], Nathan Intrator[2,3,4] and Dominique Lamy[1,4]

[1] The School of Psychological Sciences, Tel Aviv University, [2] Neurosteer Inc [3] The Blavatnik School of Computer Science, Tel-Aviv University, [4] The Sagol School of Neuroscience, Tel Aviv University

Working Memory (WM) load is an important cognitive feature that is highly correlated with mental effort. Several neurological biomarkers such as theta power and mid-frontal activity show increased activity with increasing WM load. Such correlations often break down in cognitively impaired individuals, making WM load biomarkers a valuable tool for the detection of cognitive impairment. However, most studies have used a multi-channel EEG or an fMRI, which are not massively accessible.

In the present study, we evaluate the ability of novel features extracted from a single-channel EEG located on the forehead, to serve as markers of WM load. We employed the widely used n-back task to manipulate WM load. Fourteen participants performed the n-back task while their brain activity was recorded with the Neurosteer® Aurora EEG device. The results showed that the activity of the newly introduced features increased with WM load, similar to the theta band, but exhibited higher sensitivity to finer WM load changes. These more sensitive biomarkers of WM load are a promising tool for mass screening of mild cognitive impairment.

## Introduction

Working memory (WM) is a theoretical concept which refers to a cognitive network that enables to temporarily hold a limited amount of information for immediate processing (1). The amount of information that needs to be stored in WM in order to meet task requirements at any given time, is referred to as WM load: it increases with task difficulty and is associated with mental effort (2-3). In healthy individuals, activity in a variety of brain regions and cortical oscillations were found to be consistently modulated by WM load (4-6). Such correlations often break down in cognitively impaired individuals (7-13), making WM load biomarkers a valuable tool for the detection of cognitive impairment.

The n-back task is a popular paradigm for measuring performance under varying WM loads (14-16). In a typical n-back task, participants are required to identify whether the current stimulus (generally a letter or a number) matches a stimulus presented *n* trials ago (*n* is usually 1, 2, or 3). Increasing the level of *n* renders the task more difficult, because participants need not only to retain information in WM, but also to update it continuously. Thus, the n-back task allows one to control working memory load by manipulating task difficulty, while leaving visual input and motor output unchanged.

Various studies used the n-back task to manipulate WM load during recording of brain activity with electroencephalography (EEG) in the healthy population. They found that increasing WM load is associated with theta-band power increase in the frontal regions (6,18,19) and alpha-band power decrease in the parieto-occipital midline regions (5). Similar studies with cognitively impaired populations (e.g. mild cognitive impairment (MCI) and Alzheimer's disease (AD)) revealed several EEG biomarkers of cognitive deficit severity (8-12, 17-19). For instance, anomalies in theta activity were detected as early as one year before cognitive deterioration was detected (8,9,11). Other EEG studies relying on Event-Related Potentials (ERPs) reported that the latencies of the P200 and N200 components during a two-back task were higher for participants who developed progressive cognitive decline and AD one year following the n-back task examination (9). Finally, event-related desynchronization (ERD%) on alpha and beta bands was found to be consistently lower in MCI and AD groups relative to healthy participants (10).

These findings indicate that EEG-based WM-load biomarkers are useful diagnostic tools. However, they rely on high-density multi-channel EEG systems (typically with more than 16 electrodes). Such tools are expensive, bulky and require long setup and trained medical professionals. Mobile and affordable acquisition systems may provide an appealing solution to these problems.

Two studies using such mobile systems with the n-back task identified biomarkers of WM load using EEG (20) and Functional Near-Infrared Spectroscopy fNIRS (21). These studies further employed machine learning (ML) techniques and reported more than 80% accuracy at classifying different n-back levels. However, these studies did not address the risk of overfitting the data, which is important in order to ensure generalization in classification studies: First, their feature extraction techniques were not fully defined prior to data collection. Second, their feature classification accuracies were not tested on a new independent dataset. The purpose of the present study was therefore to examine the performance of predefined features extracted using a single-channel EEG device, as WM load biomarkers. To meet this goal, we used a wearable EEG system with a 3-electrode patch placed on the participant's forehead (Aurora by Neurosteer® Inc). The system provides 121 Brain Activity Features (BAFs) extracted via harmonic analysis (see Appendix A for full details). The BAFs extraction technique was defined on a large set of EEG recordings (dataset A). Next, the BAFs were extracted from another dataset (dataset B), which included healthy participants undergoing different cognitive tasks such as auditory detection, auditory discrimination, and resting state tasks. From dataset B, new linear and non-linear combinations of BAFs (e.g. higher-level features) were calculated using machine learning (ML) algorithms. The purpose was to discriminate between the different cognitive tasks and difficulty levels. Three of these features showed high separation abilities on dataset B: VC9, ST4, and T4. Specifically, VC9 was identified as the best discriminator between an auditory detection task and an auditory classification task, ST4 was the best discriminator between different load levels in an auditory n-back task and T4 was the best discriminator between an auditory detection task and resting state. Importantly, the data from which the BAF extraction technique was designed (dataset A), and the data from which the higher-level features were calculated (dataset B), did not include the current visuo-numeric n-back task.

In the present study, we used our single-channel EEG device to measure the brain activity of 14 new participants (not included in datasets A or B), while they performed the n-back task with four WM load levels (n=0,1,2,3). Activity of VC9, T4 and ST4 was calculated from the data, in addition to the power spectral density of classical frequency bands (delta, theta, alpha, beta and gamma). A previous study using a wireless EEG device found that differences between n-back conditions were observed in the average signal power spectrum (20), therefore we used all frequency bands rather than only theta. Finally, we compared the performance of the previously extracted features (VC9, ST4, and T4) to the performance of a feature that was trained to classify between WM load levels on the data of the current study (with a larger risk of overfitting). For this purpose, we performed a Linear Discriminant Analysis (LDA) (22) on the 121 BAFs, with different WM load levels. Figure 1 depicts a schematic flow of the data collection and analysis process (for specific technical details, see Appendix A).

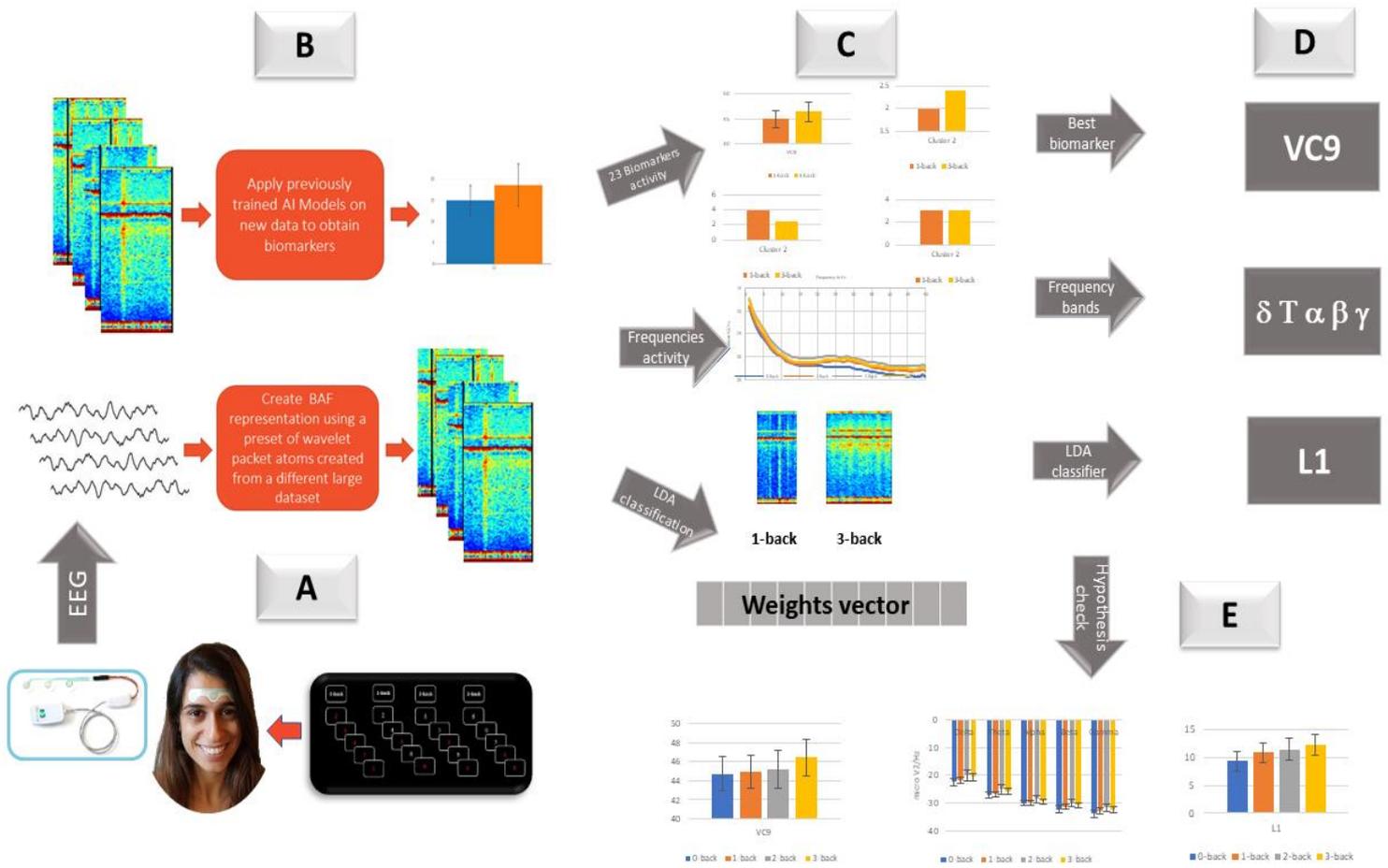

**Figure 1**. *A graphical representation of the study.*

*(A) EEG data is recorded from a single-channel using a 3-electrode forehead patch by Aurora EEG system, while participants perform the n-back task. Data is pre-processed by a predefined set of wavelet packet atoms to produce 121 Brain Activity Features (BAFs). This brain activity representation is updated once every second. From the raw EEG data, a power spectral density is produced for each of the n-back levels and an energy representation (in dB) is produced for five standard frequency bands. This representation also updates once per second.*

*(B) The BAFs are passed through several models, which were pretrained on external data, to produce 8 potential cognitive-related features. A separation between the different levels of the n-back task is sought.*

*(C) The data is additionally passed through a model trained on the study data (with potential overfitting) to produce a reference biomarker (L1) that is maximally separable from the existing BAFs.*

*(D) Three features and frequency bands are averaged across the n-back conditions. These are statistically compared to the separation power of the 5-bandwidth frequency representation and the LDA separator.*

*(E) Graphs and summary statistics of the above predictors are compared using several statistical tools.*

## Methods

*Participants*

Fourteen healthy students (8 females), mean age of 26.35 (SD=4.4), from Tel-Aviv University participated in this study for course credit. Ethical approval for this study was granted by Tel-Aviv University ethical committee on 27.3.18.

*Apparatus*

The behavioral task was administered via Lenovo Yoga 920-13IKB with Intel® Core™ i7-855OU CPU @ 1.80 GHz 1.99 GHz processor. The task was programmed and ran using MATLAB software with Psychtoolbox toolkit (6).

*EEG device and recording*

A single channel wearable EEG system was used to collect electrophysiological data (Aurora by Neurosteer® Inc). A medical-grade 3-electrode patch with dry gel for optimal signal transduction was placed on the participant's forehead. The electrodes were located at Fp1 and Fp2 and a reference electrode at Fpz. The EEG signal was amplified by a factor of a 100 and sampled at 500 Hz. Signal processing was done in the Neurosteer cloud (see Appendix A for further details). The recording room was quiet and illuminated. The researcher set up the sanitized system equipment (electrode patch, sensor, EEG monitor) and provided general instructions to the participants before starting the task. Then, the electrode patch was placed on the participant's forehead and the recording was initiated.

*Behavioral Task*

Figure 2 illustrates the n-back task. On each block, participants were presented with a sequence of digits, one at a time. At the beginning of each block, an instructions screen indicated the load factor "n" (i.e., the relative sequential position of the previous digit to which the current digit had to be compared) for the following block. The instructions screen also included a sample illustration of a digits' stream (ranging from right to left) indicating the "yes" repetitions in red color and "no" repetition in white.

Each trial began with a fixation cross that remained on the screen for either 250 or 500ms, randomly. Then, a single digit was presented for 500ms, followed by a 1500-ms blank screen inter-trial-interval (ITI). For each trial, participants responded by pressing either a "yes" key on the keyboard (D) to indicate that the current stimulus was the same as the one "n" trials ago, or a "no" key (K), to indicate that it was different. Participants were asked to respond both quickly and accurately. Participants were allowed to respond starting from the beginning of the digit presentation until the next cross presentation.

The experiment started with a 50-trial no-load block (n=0), in which the participants were asked to respond with the "yes" key to all trials. It was followed by 24 experimental blocks, each of which consisted of 30 trials: 24 (80%) of these did not include repetitions and the remaining 6 trials (20%) were repeating trials (e.g. in the 2-back condition, the current digit, n, was the same as the one that appeared before the last digit, n-2). In each experimental block, each load condition (n=1, 2, or 3) repeated exactly 8 times. Taken together, the experiment lasted 40 minutes.

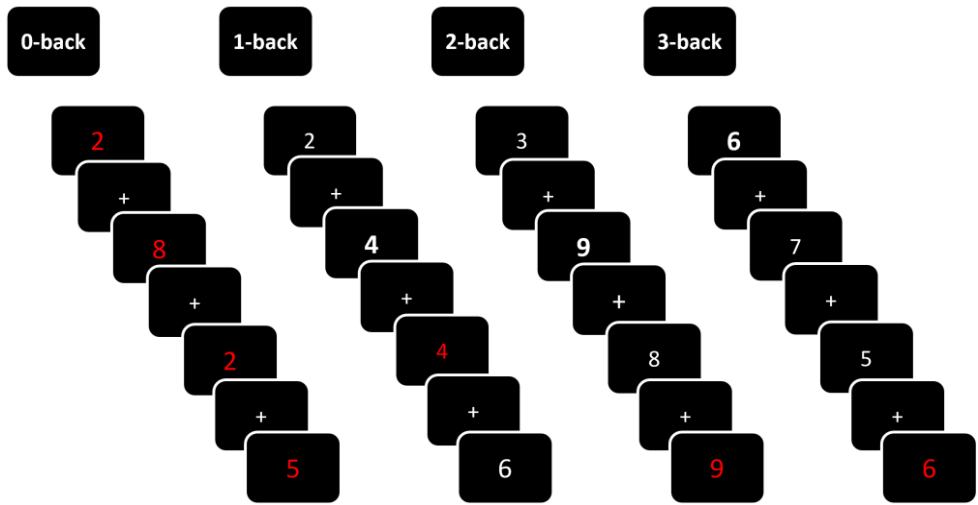

**Figure 2**. *An example of four trials in the four n-back conditions (n=0,1,2,3). For illustration purposes, white digits represent trials in which the participant was required to answer with the "No" key, and red digits represent the repeating trials, in which the participant was required to answer with the "Yes" key. Bold digits represent the digits that repeated n frames later in the block. In the actual experiment all digits had the same size and color.*

*Data Analysis*

*Construction of Brain Activity Features and Classifiers*

The full technical specifications regarding the construction of these features and the extraction of the biomarkers from the BAFs representation are provided in Appendix A. The signal processing algorithm (provided by Neurosteer) interprets the EEG data using a time/frequency wavelet-packet analysis, instead of the commonly used spectral analysis to create a new 121-dimensional feature representation. These 121 BAFs are composed from the fundamental frequencies and their high harmonics.

The BAF activity can be treated as a series of vectors with 121 values, sampled every second. Prior to the current examination, several linear and non-linear combinations of BAFs were computed using Machine Learning (ML). These combinations were extracted using data previously collected from healthy participants performing different tasks. Three of these features, VC9, ST4 and T4, were trained to classify between different levels of cognitive load and were used in the present examination. Specifically, VC9 was identified as the best separator between an auditory detection task and an auditory classification task, ST4 separated best between different load levels in an auditory n-back task, and T4 showed the best discrimination between an auditory detection task and resting state.

*Behavioral measures*

We could not apply signal detection theory (SDT, 30) to calculate sensitivity (d') because five participants had a100% hit rate on either the 1-back or the 2-back condition, and another participant had a 0% false-alarm rate on the 1-back condition. We therefore report hit rates (correct responses on repetition trials) and false alarms (incorrect responses on no-repetition trials), as well as mean reaction times (RTs) for correct responses.

*Electrophysiological variables*

The electrophysiological dependent variables included the power spectral density. Absolute power values were converted to logarithm base 10 to produce values in dB. The following frequency bands were included: delta (0.5-4 Hz), theta (4–7 Hz), alpha (8–15 Hz), beta (16–31 Hz), and lower gamma (32–45 Hz).

The BAFs analysis included the activity of the three selected features, namely, VC9, ST4 and T4, normalized to a scale of 0-100. The EEG variables were calculated per each second from a moving window of 4 seconds, and mean activity per condition was entered into the analyses.

*Linear discriminant analysis classifier*

Other than the pre-defined features mentioned above, a new variable was extracted from the data of the present study. We used a Linear Discriminant Analysis (LDA), to find the best discriminant between the 1-back and 3-back conditions, within the vector activity of the 121 BAFs. LDA is a well-known method for feature extraction and dimension reduction (22). The feature that showed the largest discrimination between the 1-back and 3-back conditions (i.e. the activity of each of the BAFs averaged across the 1-back condition and the 3-back condition) was named L1. To evaluate the discrimination abilities of this feature compared to the other EEG variables, the same statistical analyses were performed on the mean L1 activity.

**Results**

Two statistical analyses were performed on the mean of each dependent variable to determine the separation between WM load levels. The first analysis tested the difference between the no-load condition (i.e. 0-back) and load conditions (i.e. the mean of 1, 2, and 3-back) using a paired t-test. The second analysis was performed to compare between the three load conditions using a repeated-measures Analysis of Variance (ANOVA) with difficulty level of the n-back task (1, 2 or 3) as a within-participant independent variable. For significant main effects, post-hoc comparisons with Bonferroni correction of p-value for multiple comparisons were conducted. A full description of the F values, t values, p values and effect sizes ($\eta_p^2$/Cohen's D) of the t-tests and ANOVA are presented in Table 1.

*Behavioral performance*

The descriptive statistics of the behavioral measures are presented in Table 2. The difference between the no-load and load conditions was significant for all behavioral measures and so was the main effect of n-back load (1-3): reaction times (RTs) and false alarms rate were higher, and accuracy and hit rates were lower as n-back level increased. Pairwise comparisons showed that on reaction times and accuracy rates, the 1-back condition significantly differed from both the 2-back and the 3-back conditions and that on hit rates, only the difference between 1-back and 3-back was significant. The difference between 2-back and 3-back conditions was not significant for any of the behavioral measures.

| Variable | Statistic | Main effect | Post-hoc 1-back vs. 2-back | Post-hoc 2-back vs. 3-back | Post-hoc 1-back vs. 3-back | t-test 0-back vs. 1-2-3-back |
|---|---|---|---|---|---|---|
| **RTs** | F/t | 15.86 | -4.76 | -1.95 | -4.48 | -5.260 |
| | p value | **<.001** | **.001** | .222 | **.002** | **<.001** |
| | $\eta_p^2$/Cohen's d | .549 | -1.27 | -0.520 | -1.2 | 1.858 |
| **Accuracy** | F/t | 13.53 | 3.06 | 1.95 | 5.94 | 11.384 |
| | p value | **<.001** | **.028** | .122 | **<.001** | **<.001** |
| | $\eta_p^2$/Cohen's d | .510 | 0.817 | 0.52 | 1.59 | 4.456 |
| **Hit** | F/t | 11.55 | 1.52 | 2.58 | 4.78 | 9.022 |
| | p value | **<.001** | .456 | .069 | **.001** | **<.001** |
| | $\eta_p^2$/Cohen's d | .471 | 0.407 | 0.688 | 1.28 | 3.5 |
| **Delta** | F/t | 1.686 | -1.54 | 0.467 | -4.49 | -1.445 |
| | p value | .205 | .444 | 1 | **.002** | .172 |
| | $\eta_p^2$/Cohen's d | .115 | -0.411 | 0.125 | -1.2 | 0.242 |
| **Theta** | F/t | 2.62 | -1.8 | 0.56 | -6.06 | -3.073 |
| | p value | .127 | .284 | 1 | **<.001** | **.009** |
| | $\eta_p^2$/Cohen's d | .168 | -0.482 | 0.149 | -1.62 | 0.331 |
| **Alpha** | F/t | 1.33 | -1.3 | 0.843 | -3.26 | -1.880 |
| | p value | .281 | .646 | 1 | **.019** | .083 |
| | $\eta_p^2$/Cohen's d | .093 | -0.348 | 0.225 | -0.871 | 0.267 |
| **Beta** | F/t | 1.25 | -1.24 | 0.814 | -1.86 | -2.234 |
| | p value | .304 | .710 | 1 | .258 | **.044** |
| | $\eta_p^2$/Cohen's d | .088 | -0.332 | 0.217 | -0.496 | 0.346 |
| **Gamma** | F/t | 1.5 | -1.41 | 0.798 | -1.89 | -2.636 |
| | p value | .241 | .542 | 1 | .243 | **.021** |
| | $\eta_p^2$/Cohen's d | .101 | -0.378 | 0.213 | -0.506 | 0.34 |
| **VC9** | F/t | 11.46 | -3.08 | -1.72 | -4.23 | -2.255 |
| | p value | **<.001** | **.02*** | .329 | **.003** | **.042** |
| | $\eta_p^2$/Cohen's d | .469 | -0.882 | -0.459 | -1.13 | 0.234 |
| **T4** | F/t | 6.7 | 2.93 | 0.582 | 3.24 | 0.792 |
| | p value | **.005** | **.035** | 1 | **.019** | .443 |
| | $\eta_p^2$/Cohen's d | .340 | 0.782 | 0.156 | 0.866 | 0.124 |
| **ST4** | F/t | 2.5 | -1.25 | -.92 | -2.3 | -2.200 |
| | p value | .101 | .695 | 1 | .115 | **.046** |
| | $\eta_p^2$/Cohen's d | .161 | -0.335 | -0.247 | -0.616 | 0.607 |
| **L1** | F/t | 7.18 | -2.97 | 0.8 | -3.58 | -3.580 |
| | p value | **.003** | **.033** | 1 | **.01** | **.003** |
| | $\eta_p^2$/Cohen's d | .356 | -0.794 | -0.0695 | -0.956 | 0.544 |

**Table 1**. *The F, t, p values and effect sizes ($\eta_p^2$/Cohen's D) of the main effect of n-back level, post-hoc tests between the n-back levels using Bonferroni correction, and a comparison between the 0-back condition (the first block) and the other blocks of the experiment. Significant effects are marked in bold.*

|  | **0-back** | **1-back** | **2-back** | **3-back** |
|---|---|---|---|---|
| **Accuracy** (%) | 99.8±0.2 | 93.6±0.8 | 89.8±2 | 82.9±1.4 |
| **Hit rates** (%) | 99.8±0.2 | 86.6±2.2 | 83.3±3.4 | 68.7±3.4 |
| **False alarms** (%) |  | 3.31±1.7 | 7.38±2.6 | 11.4±3.1 |
| **Reaction times** (milliseconds) | 350.19±27.43 | 484.13±19.35 | 542.44±29.09 | 575.54±37.98 |

**Table 2**. *Mean values of accuracy, hit rates, false alarms and reaction times for the four n-back conditions (0, 1, 2 and 3).*

*Frequency bands*

A plot of frequency between 1-50 Hz for each n-back condition is presented in Figure 3A. The difference between the no-load conditions and load conditions was significant for the theta, beta and gamma frequency bands. The main effect of n-back level (1-3) did not reach significance for any of the frequency bands (for the averaged activity per frequency band, see figure 3B) after degrees of freedom correction because of unequal variance of the 2-back condition. The difference between the 1-back and 3-back conditions was significant for theta, delta, and alpha bands. The differences between intermediate levels (1 vs. 2 and 2 vs. 3) was not significant for any of the frequency bands.

A

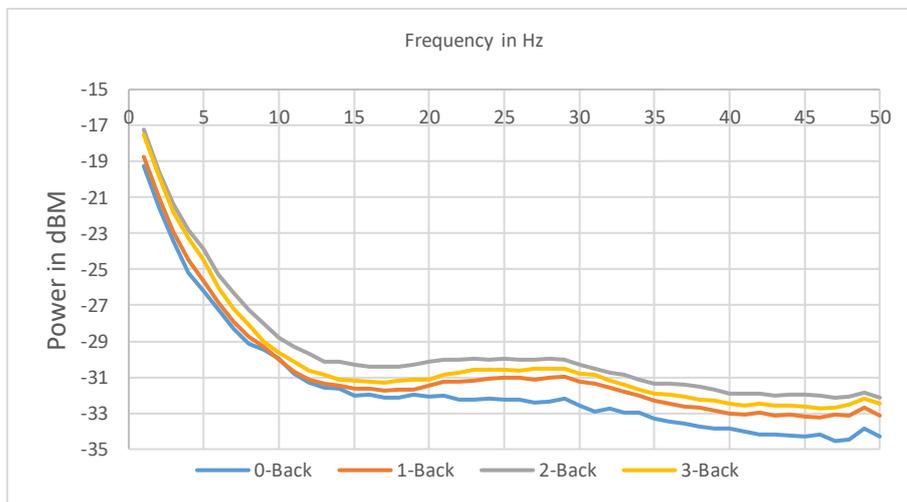

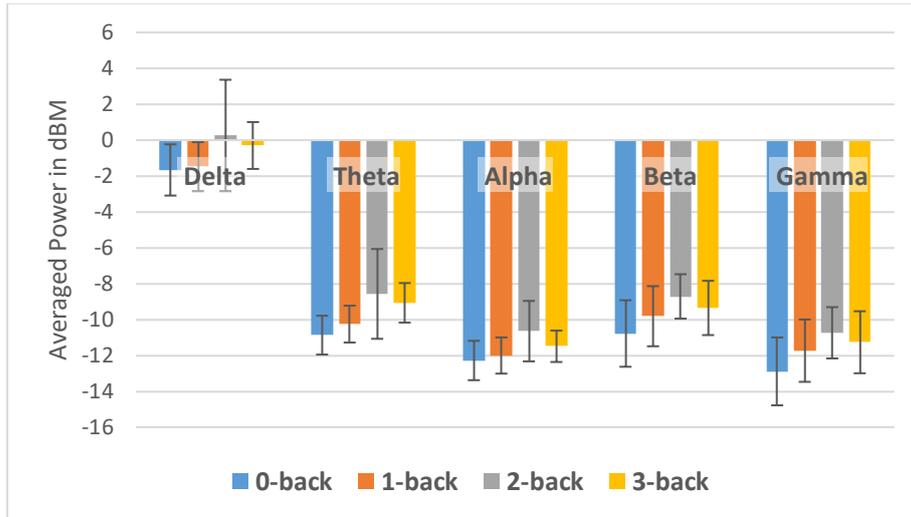

**Figure 3**. *Mean power (micro $V^2$/Hz) of each frequency between 1-50 hz (A), and of the delta, theta, alpha, beta and gamma frequency bands (B), for the 0-back (blue), 1-back (red), 2-back (grey) and 3-back (yellow) conditions. Error bars represent standard errors with Morey's correction (20).*

*Features*

A summary of the VC9, T4 and ST4 results is presented in Figure 4. The difference between the no-load and load conditions was significant for VC9 and ST4. The main effect of n-back load (1-3) was significant for VC9 and T4 and so were the differences between the 1- vs. 2- back and the 1- vs. 3-back conditions. The difference between the 2- and 3-back conditions was not significant for any of the features.

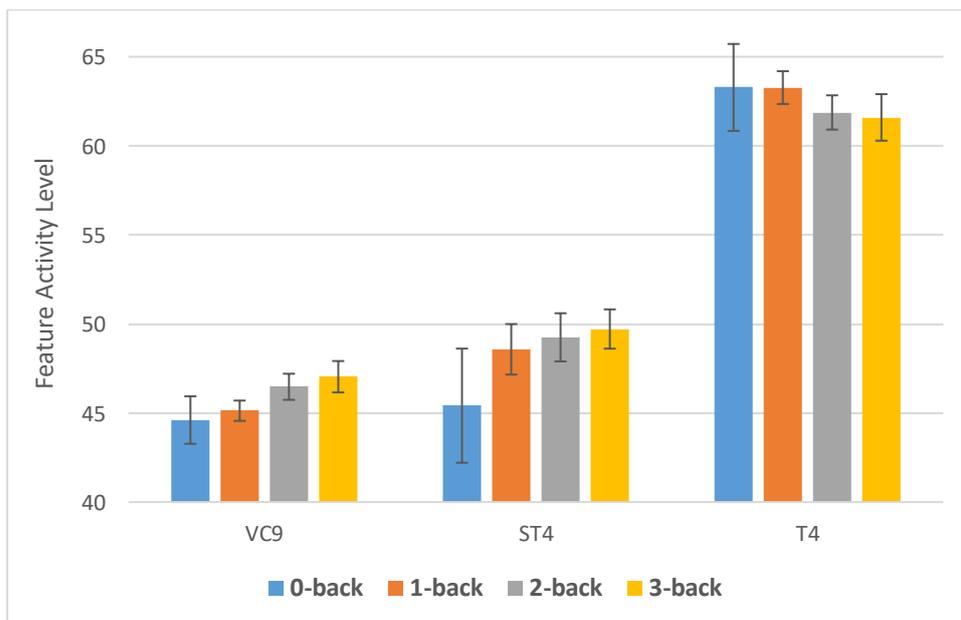

**Figure 4**. *Mean power of VC9 ST4 and T4 features (normalized between 0-100) during the n-back task, for the 0-back (blue), 1-back (red), 2-back (grey) and 3-back (yellow) conditions. Error bars represent standard errors with Morey's correction (20).*

*Linear discriminant analysis classifier*

Linear Discriminant Analysis (LDA) (22) was used to find the transformation of the 121 BAFs that would produce maximal separation between the 1-back and 3-back conditions. This resulted in a 121-weight vector applied to each second of the 121 BAFs activity, producing a new feature named L1. The mean activity of L1 is presented in Figure 5 (see Table 1 for statistics). The difference between the load and no-load conditions was significant, as well as the main effect of n-back load. The critical difference between the 1-back and 3-back conditions was also significant. Finally, Pearson correlation between the L1 and RT individual slopes revealed a non-significant correlation (Pearson R=0.441, $p=.115$).

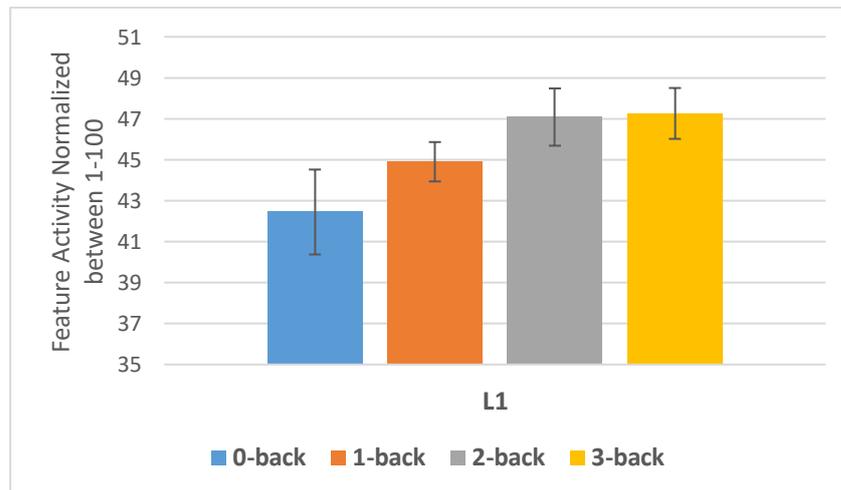

**Figure 5**. *Mean power of the L1 feature (normalized between 0-100) during the n-back task, for the 0-back (blue), 1-back condition (red), 2-back (grey) and 3-back (yellow) conditions. Error bars represent standard errors with Morey's correction (20).*

*Exploratory analyses*

*Correlations between individual behavioral performance and EEG variables*

Following Rympa et al. (4) we calculated the individual RT slope (the least-squares estimated coefficient of a linear regression of RT on memory load interpolated between 1-back and 3-back conditions; see figure 6b). These varied considerably, from -11 for participant AY to 132 for participant AG. The same slope between the 1-back and 3-back conditions was also calculated for each of the EEG variables (i.e., both frequency bands and features).

To test the relationship between the EEG activity and individual performance, we calculated the Pearson correlation between individual RT slope and EEG variables slopes. This correlation was found to be significant only for the ST4 feature, Pearson R= 0.58, *p=.03* (see Table 3).

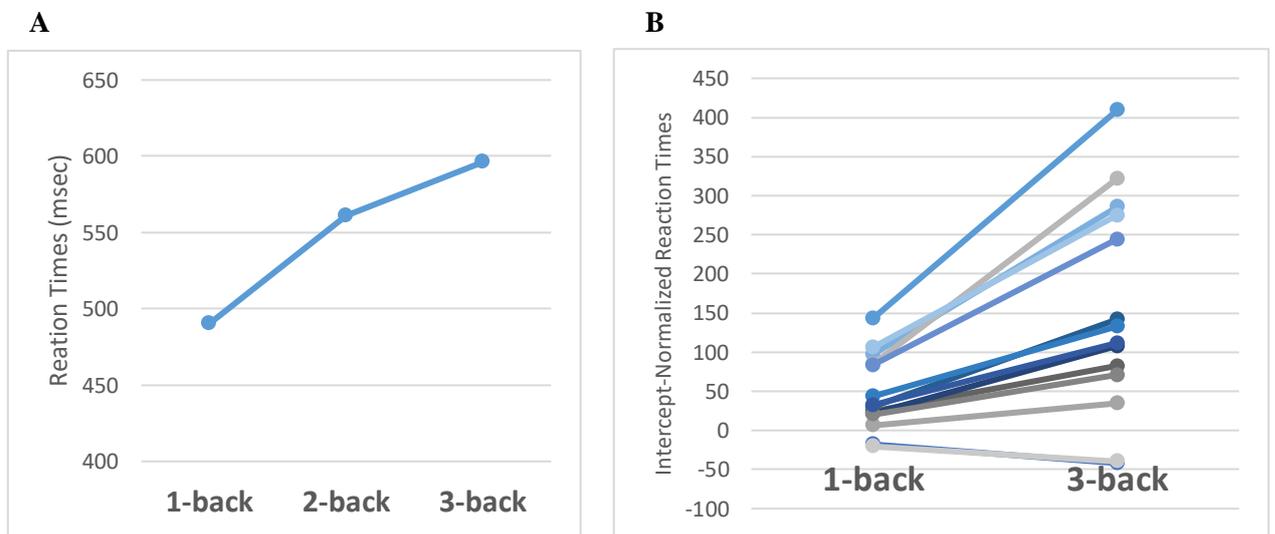

**Figure 6**. *(A) Mean reaction times by load condition (1-back, 2-back and 3-back), (B) Intercept-normalized individual RT slopes.*

| Correlation parameter | Delta slope | Theta slope | Alpha slope | Beta slope | Gamma slope | VC9 slope | T4 slope | ST4 slope |
|---|---|---|---|---|---|---|---|---|
| RT slope Correlation Coefficient Pearson R | 0.356 | 0.008 | -0.296 | -0.328 | -0.353 | 0.295 | -0.059 | 0.578 |
| Sig. level (2-tailed) | .212 | .978 | .304 | .253 | .215 | .306 | .840 | **.030** |

**Table 3**. *Summary of correlation coefficients and p-values for the correlation between the individual RTs slope and each EEG variable slope. Significant effects are in bold.*

**Discussion**

This study examined whether measures derived from a single-channel EEG device could serve as WM load biomarkers. To that end, we administered the widely used n-back task to fourteen healthy young participants. As expected, the results showed that behavioral performance decreased as WM load increased.

Using a single-channel EEG device, we extracted classical frequency bands (delta, theta, alpha, beta, and gamma), a novel feature L1 computed from the same data set (potentially overfitting the data) and novel features VC9, T4, ST4 computed from an independent dataset. We evaluated the performance of each of these measures at discriminating between different WM load levels. In line with previous studies (6,8,19), the theta band provided good discrimination between extreme levels of WM load (no-load vs. load and 1- vs. 3-back), while other classical frequency bands failed to differentiate either between the no-load vs. load conditions (delta and alpha) or between the 1- vs. 3-back conditions (beta and gamma). Crucially, none of these measures discriminated between subtler variations in WM (i.e. 1- vs. 2-back or 2- vs. 3-back). By contrast, we found that VC9 and L1 outperformed the theta band, as they were also sensitive to a subtler variation in WM load (1- vs 2-back). Notably, VC9 (which was extracted from an independent dataset), was as sensitive to this variation as L1 (which was extracted from the same dataset). T4 also emerged as a sensitive biomarker of WM load, as it discriminated between the 1- vs. 3-back conditions and between the 1- vs. 2-back conditions (but it failed to discriminate between the no-load and load conditions). Finally, based on previous literature linking WM-load related RT slopes and pre-frontal activity (19), we conducted exploratory correlation analyses between individual slopes (between 1-back and 3-back) on RT and EEG. Out of the 9 EEG measures used here, only ST4 slopes showed a significant correlation with RT slopes. However, this post-hoc finding requires a replication with a larger sample.

The higher sensitivity of VC9 (and to some extent also of T4) relative to the theta band is a particularly prominent advantage of the present method for diagnostic purposes. Indeed, it is often impossible to administer high-load conditions of the n-back task (i.e. 3-back) to clinical populations, because these participants are unable to perform the task at such difficulty level (8-11). Therefore, the finding that VC9 is sensitive to differences at mid-WM load levels sets it apart as a promising diagnostic tool.

Although the present study was limited in scope (14 participants, all young and healthy students), it shows that an easy-to-use and affordable single-channel EEG can provide a tool for extracting sensitive and reliable biomarkers of WM load. If the present findings are validated by further research on cognitively impaired populations, the system and signal processing algorithms presented here promises to considerably expand the accessibility of cognitive decline evaluation.

# Appendix A: Methodological details

The data analysis methodology is composed of three steps:
1. Creation of a brain activity representation by novel Brain Activity Features (BAFs)
2. Creation of Novel Biomarkers based on the BAFs
3. Examination of the features on previously unseen data

Each of the steps is described below.

*Creation of Brain Activity features (BAFs)*

The creation of the Brain Activity Features (BAFs) occurs prior to application of the methodology onto the new data to be analyzed. Calculation of the BAFs is based on collecting a large cohort of high dynamic amplitude and frequency range single channel EEG data. The cohort includes multiple subjects that are exposed to different cognitive, emotional, and resting tasks. A schematic representation of the signal processing is depicted in Fig A1. The signal processing module is decomposing the EEG signal input into a large number of components which comprise the Brain Activity Features (BAFs). The output of the module is a Brain Activity Representation which is constructed based on the BAFs for any given EEG signal.

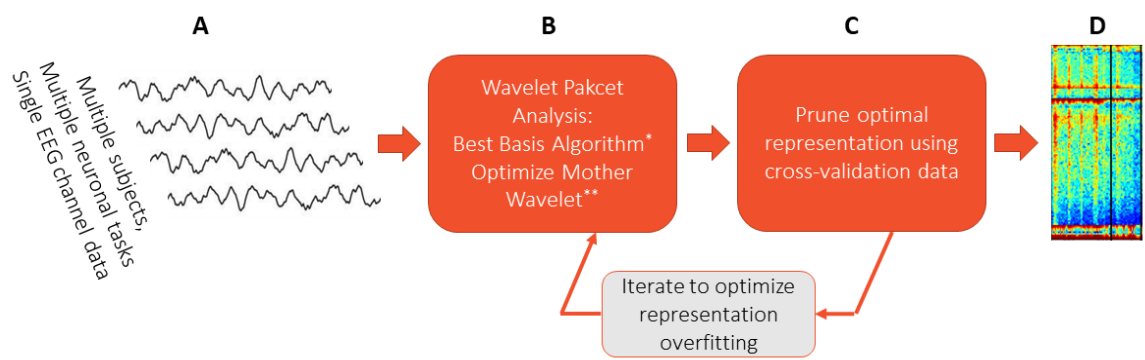

Figure A1. schematic representation of the construction of the Brain Activity Features (BAFs). See text for the description of the different steps.

**A: electrophysiological signal input**

The EEG cohort described above is the input of the signal processing algorithm presented as the first step of the process.

## B: Wavelet Packet Analysis

For a given cohort of EEG recordings, a family of *wavelet packet trees* is created. For the mathematical description, we follow the notation and construction provided in chapters 5, 6 and 7 of Wickerhauser's book[1].

To demonstrate the process; let *g* and *h* be a set of *biorthogonal quadrature filters* created from the filters *G* and *H* respectively. Each of these is a convolution-decimation operator, where in the case of the simple *Haar* wavelet, *g* is a set of averages and *h* is a set of differences.

The construction of the full wavelet packet tree is by successive application of these functions (Figure A2), so that at every level, a new full orthogonal decomposition of the original signal *x* is created. In the classical wavelet decomposition by Daubechies[2], only the marked parts are used and the signal is decomposed into *Gx, GHx etc.,* but the full construction of the tree continues recursively, on *Gx, GHx* and so forth, to create a full binary tree. Coifman and Wickerhauser[3] observed that a large number of orthogonal decompositions can be constructed from the full tree by mixing between the different levels and different blocks of the tree, following a simple rule. The recursive construction of the full tree is described next.

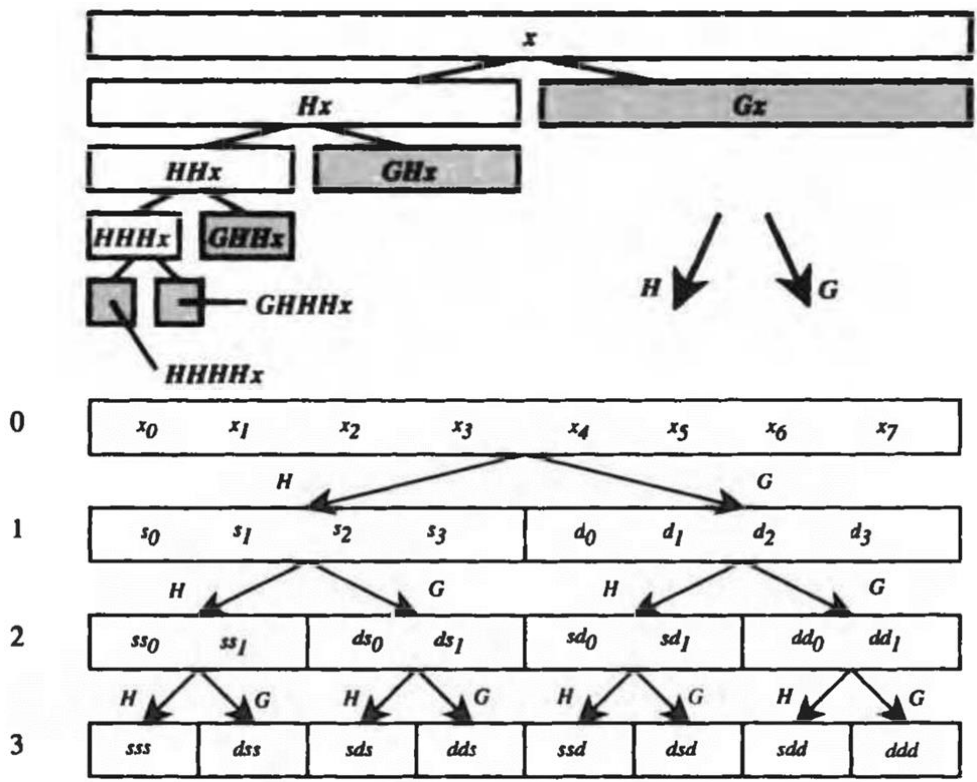

Figure A2. Construction of a Discrete Wavelet Transform Tree (Taken from Wickerhauser[1]). The top panel represents the classical wavelet construction and the bottom panel extends the construction to a full wavelet packet tree.

Let $\psi_1$ be the *mother wavelet* associated to the filters $s \in H$, an $d \in G$. Then, the collection of *wavelet packets* $\psi_n$, is given by:

$$\psi_{2n} = H\psi_n; \quad \psi_{2n}(t) = \sqrt{2} \sum_{j \in Z} s(j)\psi_n(2t - j),$$

$$\psi_{2n+1} = G\psi_n; \quad \psi_{2n+1}(t) = \sqrt{2} \sum_{j \in Z} d(j)\psi_n(2t - j).$$

The recursive form provides a natural arrangement in the form of a binary tree (Figure A2). The functions $\psi_n$ have a fixed scale. A library of wavelet packets of any scale *s*, frequency *f,* and position *p* is given by:

$$\psi_{sfp}(t) = 2^{-s/2}\psi_f(2^{-s}t - p).$$

The wavelet packets $\{\psi_{sfp}: p \in Z\}$ are an orthonormal basis for every *f* (under orthogonality condition of the filters *H* and *G*) and are called *orthonomal wavelet packets*.

Using this construction, Coifman and Wickerhauser applied the *best basis* algorithm[3] to search for an orthonormal base that satisfies a specific optimality condition. The optimality condition that was chosen is Shannon's entropy of the coefficients of each component (or wavelet packet atom). It is a measure that prefers coefficients with a distribution that is far from uniform, in the sense that it prefers a distribution with a small number of high value coefficients and a long tale, namely, a large number with low value coefficients. The full details of the best basis search are described in chapter 7 of Wickerhaser's book.

The process of creating a best basis from the wavelet packet tree can be further iterated by an optimization on the mother wavelet using a gradient descent in wavelet space as is described in Neretti and Intrator[4].

**C: Pruning the optimal representation**

The outcome of the best basis algorithm is an orthogonal decomposition that is adapted to the stochastic properties of the collection of EEG signals. However, there is a risk that the decomposition is "overfitting" namely it is too adapted to the EEG signals from which it was created. To avoid this phenomenon, we first have to get rid of "small" coefficients. This can be done by the denoising technique of Coifman and Donoho[5]. The next step is introducing a validation set, which is another collection of EEG-recordings that was not used in the creation of the best basis. Using this set, we can determine which atoms maintain a high energy (some large coefficients) when decomposing the new signals. These atoms will remain in the representation. At the end of this part, the resulting set of decomposing signal contains only a part of the full orthonormal basis that was found. We then reorder the basis components not based on the binary tree that created them, but based on the correlation between the different components In this way, we created a brain activity representation in which components that are more correlated to each other, are also geographically close to each other within the representation. This is done for the purpose of improved visualization.

**D: brain activity representation output**

The result of the signal processing module is the brain activity representation. Specifically, it is a collection of 121 energy components, emanating from the wavelet packets as well as standard frequency bands which are updated each second. The representation (D) shows a color heatmap of each of the 121 X time matrix, so that the x axis represents time and the y axis represents the different components.

*Creation of Novel features based on the BAFs*

The signal components, which we termed BAFs, were constructed from single EEG channel recordings in an unsupervised manner, namely, there were no labels attached to the recordings for the purpose of creating the decomposition. To create biomarkers based on the BAFs, task labels are used, indicating the nature of cognitive, emotional, or resting challenge the subject is exposed to during the recording.

Given labels from a collection of subjects, and the corresponding high-dimensional BAF data, a collection of models attempting to differentiate between the labels based on the BAF activity can be used. In the linear case, these models are of the form:

$$V_k(w,x) = \Psi\left(\sum_i w_i x_i\right),$$

where w is a vector of weights, and $\Psi$ is a transfer function that can either be linear, e.g., $\Psi(y) = y$, or sigmoidal for logistic regression $\Psi(y) = 1/(1 + e^{-y})$.

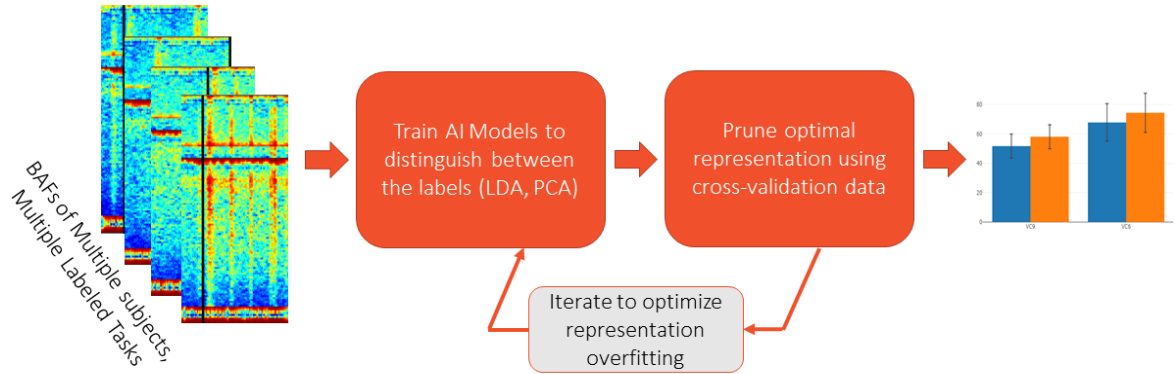

Figure A3. Supervised construction of different features from labeled brain activity representation of different cognitive and non-cognitive tasks.

For each predictor, which we term biomarker, a standard machine learning procedure is applied as follows:

1. Choose a labeled data set with at least two different tasks (e.g. cognitive, emotional, or resting challenge). The data set may include the same challenge but for a non-homogenous group.
2. Separate the data into three sets: training, validation, and test.
3. Choose a model to train on from a family of models that includes linear regression, linear regression with binary constrains (zero and one values for the weights), linear regression with only positive values, logistic regression, discriminant analysis and principal components analysis. In the non-linear models, use neural networks, support vector machine and the like.
4. Train each model on several sets of train/test and validation to best estimate internal model such as the variance constraints, on the ridge regression, the kernel size and number of kernels in a support vector machine, or the weight constraints in a neural network model.
5. From the above models, obtain predictors to be tested on other data with potentially other cognitive, emotional and rest challenges.

6. The last step in the process includes testing the biomarkers using a test data labeled set that was not used in the creation of these features. This allows removal of features that were overfitting to the training data, namely, they do not produce high significant difference on the validation data. This is still part of the model creation and not part of the model testing that is done on new data and described in step 3.

All above steps are described in the scheme on Figure A3.

*Examination of the features on previously unseen data*

Following the creation of BAFs and the creation of features as described above, the features relevance can be tested on various cognitive or emotional challenge. The testing scheme is described in Figure A4. Specifically, data is collected with the sensor system and sent to the cloud for creation of a BAF representation using the previously determined wavelet packet atoms. The BAF representation is provided to previously determined ML models, which convert the BAF activity into features. Statistical tests are then applied to determine the quality of the predictions and the correlation of the features to the cognitive and emotional challenges that the participants undergo. This may include single subject analysis as well as group analysis.

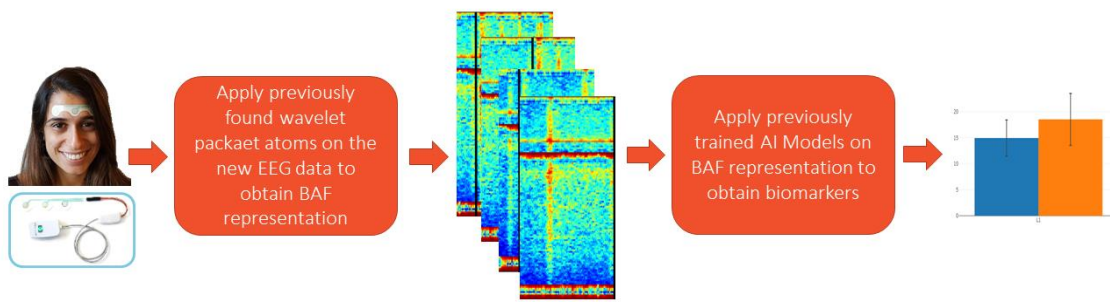

Figure A4: Testing the relevance of the previously found features on the data.

In the process of testing the features on new data, we may want to get an *upper bound* to the performance of the feature, by seeking an *overfitting biomarker* on the currently tested data. This is only done to get an idea of the potential upper bound on prediction abilities from the existing data, and indirectly can tell us more about the optimality of the actual features that were constructed from a different data set and are assumed to be more general in this sense.

# Appendix References